\documentclass[%
 reprint,
superscriptaddress,
 amsmath,amssymb,
 nobalancelastpage,
 aps,
prb,
]{revtex4-1}
\usepackage[english]{babel}
\usepackage{combelow}
\usepackage{color}
\usepackage{multirow,cprotect}
\usepackage[justification=raggedright,singlelinecheck=false]{caption}
\usepackage{subcaption}
\usepackage{graphicx}
\usepackage{dcolumn}

\usepackage{bm}
\usepackage{amsmath}

\begin{document}

\preprint{APS/123-QED}

\title{Characterization of the BIFROST spectrometer through virtual experiments}

\author{Kristine M. L. Krighaar}
 \email{kristine.krighaar@nbi.ku.dk}
\affiliation{Nanoscience Center, Niels Bohr Institute, University of Copenhagen, 2100 Copenhagen, Denmark}

\author{Silas B. Schack}
\affiliation{Nanoscience Center, Niels Bohr Institute, University of Copenhagen, 2100 Copenhagen, Denmark}

\author{Nicolai L. Amin}
\affiliation{Technical University of Denmark, Lyngby Denmark}
\affiliation{European Spallation Source ERIC, Data Management \& Scientific Computing Division, Asmussens All\'e 305,  2800 Lyngby, Denmark}%

\author{Gregory S. Tucker}%
\affiliation{European Spallation Source ERIC, Data Management \& Scientific Computing Division, Asmussens All\'e 305,  2800 Lyngby, Denmark}%

\author{Rasmus Toft-Petersen}%
\affiliation{Technical University of Denmark, Lyngby Denmark}%
\affiliation{European Spallation Source ERIC, Lund, Sweden}%

\author{Kim Lefmann}%
\email{lefmann@nbi.ku.dk}
\affiliation{Nanoscience Center, Niels Bohr Institute, University of Copenhagen, 2100 Copenhagen, Denmark}%

\date{\today}

\begin{abstract}

Using the Monte Carlo ray tracing package McStas, we illustrate the possibilities of creating virtual experiments of the neutron spectrometer BIFROST at the European Spallation Source, ESS. With this model, we are able to benchmark BIFROST with respect to expected intensity, $Q$- and energy-resolution. The simulations reproduce the expected resolution behavior and quantify effects that are difficult to capture analytically, including a wavelength-dependent edge enhancement arising from a combination of the long-pulsed source and the pulse-shaping chopper. Furthermore, we present an antiferromagnetic (AF) spin wave simulation, which we use to create realistic datasets at different instrument operation settings. Our virtual experiments focus on realistic dispersive dynamics and illustrate how the virtual experiment approach reveal resolution effects, not easily calculable via analytical models. This demonstrates the crucial role of numerical simulations in the planning of challenging experiments.
\end{abstract}

\pacs{74.72.-h,75.25.-j,75.40.Gb,78.70.Nx}

\maketitle

\section{Introduction}

Among the 15 planned instruments at the European Spallation Source (ESS), five will be spectrometers specialized for inelastic neutron scattering\cite{andersen_instrument_2020}.
In this type of scattering, the measured intensity is proportional to the dynamical correlation function, $S({\hbar Q}, \Delta E)$ that can be calculated by theoretical models \cite{boothroyd_principles_2020}. Here, $\Delta E$ is the neutron energy transfer and $\hbar Q$ is the momentum transfer.

Spectroscopy with cold neutrons is in particular used for high-resolution studies of magnetic excitations in quantum materials for values of $\Delta E$ between 0 and 10~meV and a range in $Q$ between $\sim 0.5$ and 3.0~\AA$^{-1}$. Systems of current interest include quantum magnets \cite{Coldea2010, Mourigal2013, IN14_1, LET14, MACS9, AMATERAS21, MACS21, Thales25, CNCS63}, frustrated magnets \cite{Han2012, LET6, IN5_15, Fennell2019, OSIRIS6, IN5_34, AMATERAS18, DCS28, CNCS45, Thales27}, multiferroics \cite{fabreges_spin-lattice_2009, PANDA2, Ratcliff2016, Bao2023},  van der Waals magnets \cite{HYSPEC7, SIKA5}, and superconductors \cite{Christianson2008,IN12_1, Wang2016,Thales5,CNCS31,LET20}.

\begin{figure*}
    \centering
    \includegraphics[width=\linewidth]{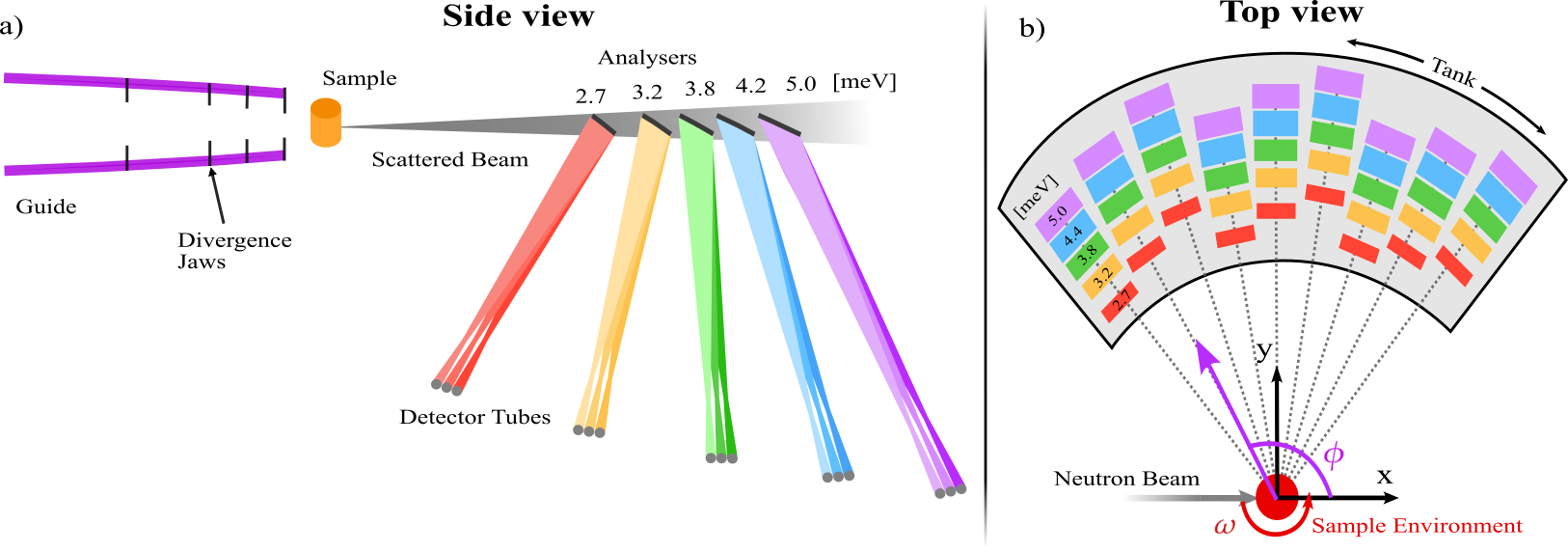}
    \caption{(a) Side view sketch of the last meters of the BIFROST spectrometer, including guide end, divergence jaws, sample, and the secondary spectrometer (analysers and detectors) (b) Top view of the secondary spectrometer showing the sample (orange) and the 45 analyser assemblies (blue). This subfigure is modified from \cite{toft-petersen_bifrostindirect_2025}.}
    \label{fig:lay-out}
\end{figure*}

One of the ESS instruments specialized for cold neutron spectroscopy is the indirect geometry spectrometer BIFROST \cite{toft-petersen_bifrostindirect_2025}, positioned at the end of a 160~m long guide between the moderator and the sample. The design of BIFROST is an enhancement of the classical indirect spectrometer design by having 5 rows of PG analysers behind each other. Each analyser bank reflects a different energy band onto a triplet of detectors, using the prismatic concept to improve energy resolution \cite{birk_prismatic_2014}. Neutrons that do not satisfy the given Bragg condition are transmitted to the next bank of analyzers. The lay-out of the BIFROST secondary spectrometer is very similar to the monochromator-type spectrometer CAMEA at PSI \cite{lass_commissioning_2023} and is sketched in Fig.~\ref{fig:lay-out}. 

It is common to perform analytical estimations of neutron instrument performance, and simulation tools are widely used as a highly valuable and popular means of testing these expected behaviours. While this approach provides accurate descriptions of certain aspects, such as the elastic energy resolution, other regions of $(Q,\Delta E)$ remain less explored. Furthermore, during commissioning and experiments, resolution-related effects may emerge that are not fully captured by these models because of the inherent complexity of real instruments. These challenges highlight the ongoing need to further develop both theoretical and simulation approaches, enabling more accurate predictions of instrument performance. We aim for the framework presented here to both address these challenges and to inspire and help establish new standards for performing neutron instrument simulations.

In this work, we present a detailed simulation of the BIFROST instrument using the McStas package, with the aim of showing the benefits of predicting instrument performance through virtual experiments \cite{lefmann_virtual_2008}. We use well established analytical approaches for estimating the energy resolution for BIFROST, combining analytical calculations and simulation techniques. We then extend this analysis to examine the influence on the energy resolution from the edges of the wavelength band. In addition, we study the effect of different instrument configurations on the $Q$-resolution. Building on these understandings of the $Q$- and energy-resolution, we present a detailed view of their interplay in a more realistic commissioning scenario by modelling a dispersive magnon mode in the physical system MnF$_2$ both with and without applied magnetic field. This comprehensive approach provides a deeper understanding of the capabilities and limits of the instrument, which will be essential when planning future experiments and interpreting their data.

\section{Model setup}

We employ the Monte Carlo ray-tracing software McStas\cite{lefmann_mcstas_1999,willendrup_mcstas_2020} to perform the simulations. To realistically obtain an inelastic neutron signal from BIFROST, the McStas instrument contains all the core components. The simulation of the spectrometer front end (moderator, guide, choppers, and slits) was described elsewhere \cite{holm-dahlin_optimization_2019} and is not elaborated here. However, it should be mentioned that the beam pulse is shaped by a set of {\em pulse-shaping choppers} (PSC) 6.35~m from the moderator. The opening time of these choppers can be adjusted continuously between fully open and 0.1~ms. In addition, the {\em bandwidth choppers} (BWC) select from the full spectrum a wavelength band of width 1.76~\AA.  
Finally, the divergence of the beam reaching the sample can be modified from 4 sets of {\em divergence jaws} in the final part of the elliptically focusing guide before the sample.\cite{toft-petersen_bifrostindirect_2025}

For the simulations in this work, the beam from the primary spectrometer, up until the end of the guide, was precomputed and stored into an MCPL-file\cite{kittelmann_monte_2017}. In turn, this file was used as a virtual source for the simulations of the sample and secondary spectrometer. This significantly improved efficiency by reusing neutron rays that reach the sample position. A minor smearing was applied to this virtual source to avoid repetition artifacts\cite{mcstasorg_mcpl_input_2025}; however, this did not affect the instrument resolution.

\begin{figure}
    \centering
    \includegraphics[width=1.0\linewidth]{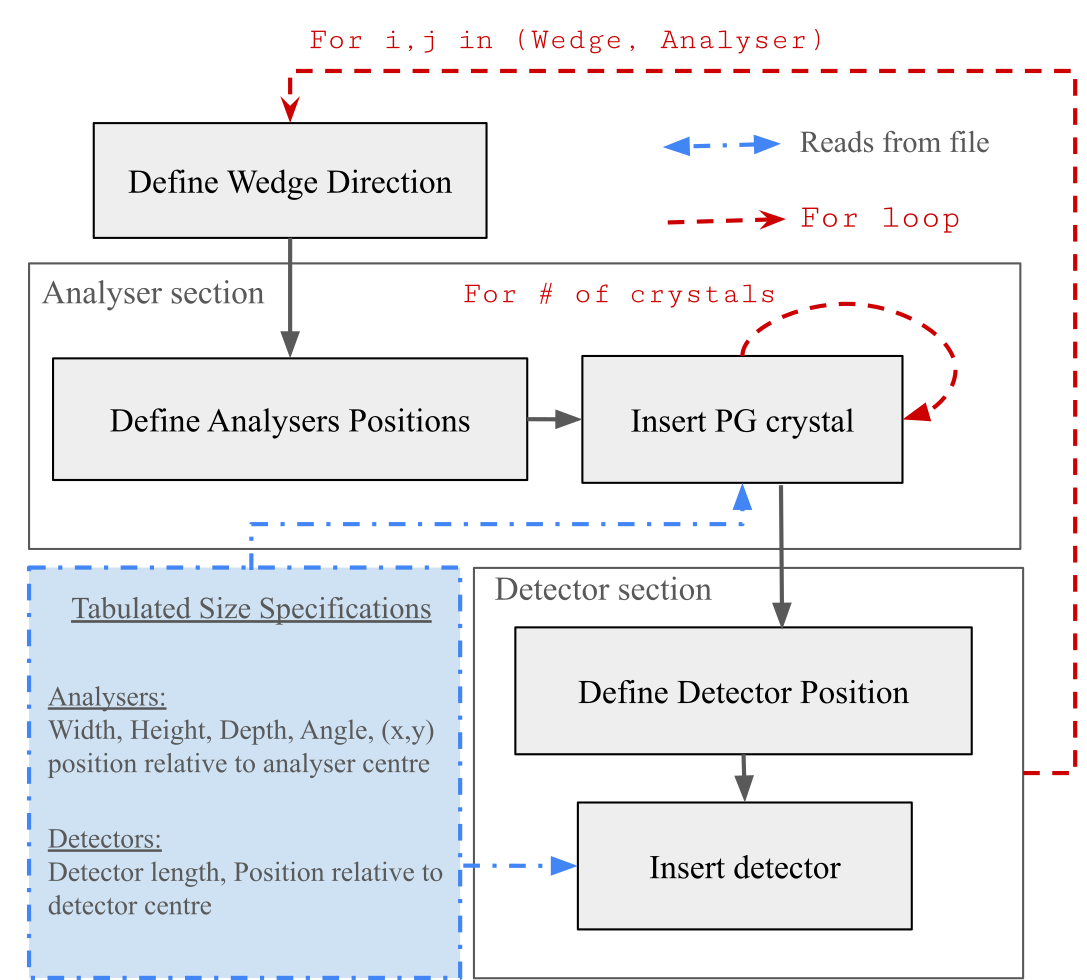}
    \caption{Flowchart illustrating the overall workflow of the BIFROST McStasScript code. This building structure allows creating any or all analyser/detector pairs in the secondary spectrometer.}
    \label{fig:McStas_flowchart}
\end{figure}

The simulation set-up of the secondary spectrometer geometry is more elaborate. Due to the computational cost of simulating the complete analyser and detector assembly, a modular approach was adopted, utilizing the python API to McStas, McStasScript\cite{bertelsen_welcome_nodate}. This allows individual sections, such as wedges or analyser banks, to be enabled or disabled as needed. A backend-building function, illustrated by Fig.~\ref{fig:McStas_flowchart}, was used to automatically generate analyser–detector pairs based on configuration tables. For the present study, only one analyser energy ($E_f = 5.0$~meV) and a single detector tube per triplet (center tube) were included to reduce simulation time. The sample stage was defined with a rotation axis similar to that of the real instrument, allowing realistic alignment within the instrument coordinate system. While the real instrument utilizes event mode formation, the presented version utilizes histogramming mode, to leverage computational efficiencies inherent to the McStas runtime.\newline  
 
For the data reduction, a data object class has been defined, as it allows for systematic organization of the data and of the meta-data that are acquired in each measurement. This python class is called \texttt{BIFROST\_measurement}, and the class converts the McStas output into a \texttt{txt}-list file with the values of $q_h$, $q_k$, $\Delta E$, $I$ and $\delta I$ for the individual pixels. 

The nominal energy transfer, $\Delta E$, is calculated using  the measured time-of-flight, ToF, and the (calibrated) final neutron energy of the particular analyser-detector pair. In this way, we determine the energy transfer

\begin{equation}
    \Delta E = E_{\rm i} - E_{\rm f} = \frac{L_{\rm i}^2 m_n}{2  (t_{\rm ToF} - t_{\rm f})^2} - E_{\rm f} .
    \label{eq:energy_transfer_bifrost}
\end{equation} 

Here, $t_{\rm f} = \sqrt{\frac{L_{\rm f} ^2 m_n}{2 E_{\rm f}}}$ is the travel time in the secondary spectrometer, $m_n$ is the neutron mass, $L_{\rm f}$ is the secondary spectrometer length from the sample to the detector, and $E_{\rm f}$ is the calibrated analyser energy. $t_{\rm ToF}$ is the time of flight measured for the neutron in the detector, $L_{\rm i}$ is the primary spectrometer flight path. In case the chopper is open ($>2.3$~ms) both $L_{\rm i}$ and $t_{\rm ToF}$ is calculated from the moderator instead of the PSC. 

For binning and reshaping the dimensions of the data, the analysis package \textit{scipp} is utilized, with the exact purpose of providing a user-friendly package for handling neutron instrument data \cite{heybrock_scipp_2020}. One of the strong benefits of scipp is that it allows for great flexibility of multidimensional data while still handling the correct labels across all dimensions.

All the necessary code to replicate the results presented in this work can be found readily available on github\cite{krighaar_github_2025}.

\section{Energy resolution}
\label{sect:energy_resolution}

As highlighted in the introduction, one of the central questions that can be addressed using McStas models is the behavior of the instrument energy resolution across different operating conditions. Instrumental broadening originates from finite distributions in neutron parameters, whose standard deviations propagate through the measurement process.  
Error propagation applied to the energy-transfer 
eq.~(\ref{eq:energy_transfer_bifrost}) has been documented extensively for other indirect time-of-flight instruments \cite{ telling_spectroscopic_2005,bewley_faro_2019,perrichon_spectral_2022, winn_flexible_2022}. However, instruments are designed differently and have different use specifications, {\em e.g.}\ high vs.\ low $\Delta E$. Instruments may also have different leading terms in the energy resolution. Following the established treatments, we identify and discuss the terms most relevant for the performance of BIFROST.

\begin{figure}
    \centering
    \includegraphics[width=\linewidth]{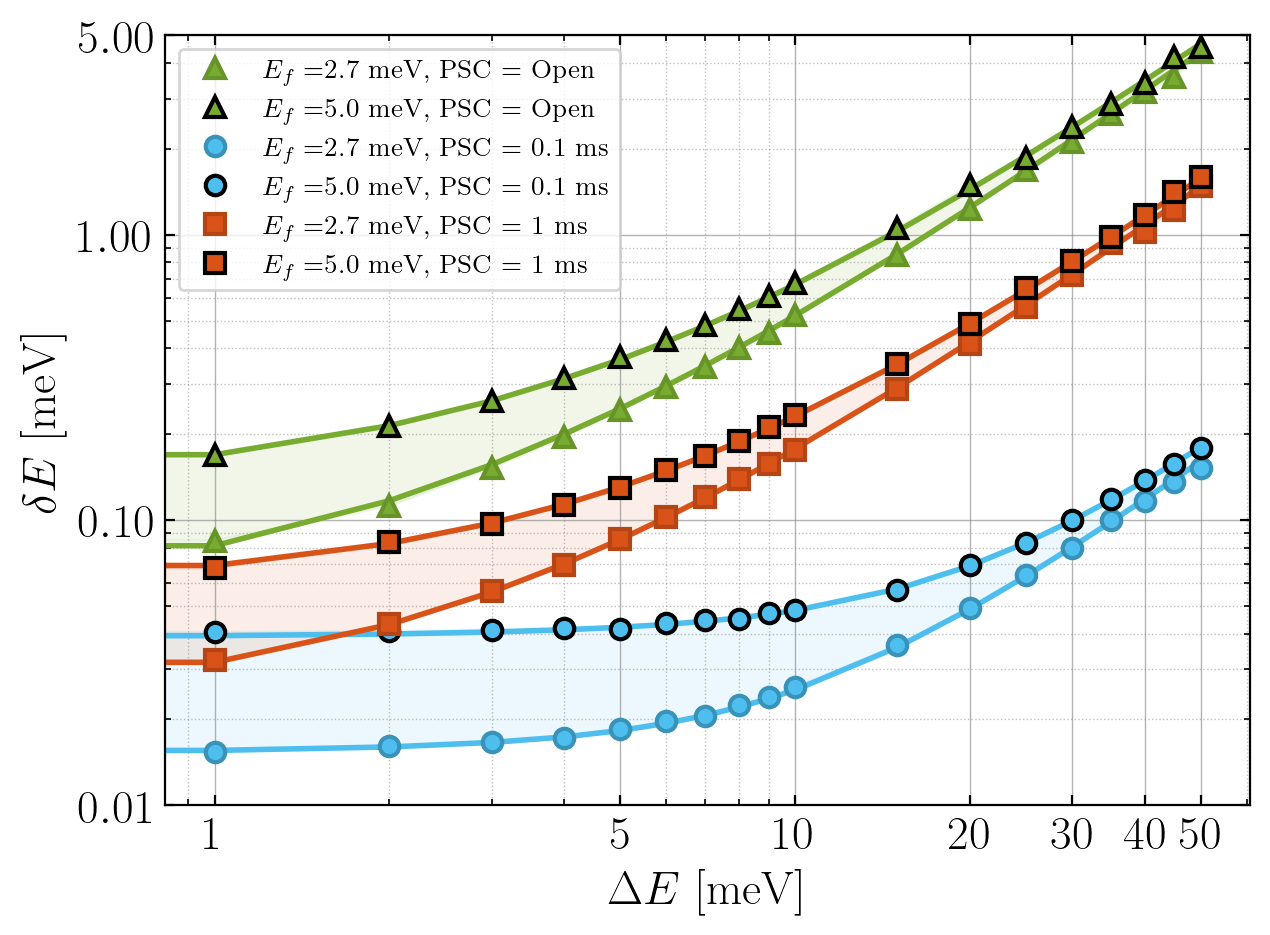}
    \caption{Comparison between simulated and analytically calculated energy resolutions of BIFROST; both given as FWHM. Markers indicate simulations result at PSC opening times of 0.1~ms, 1~ms and fully open. The black outlined markers are for the 5~meV analyser bank while the colored outline is for the 2.7~meV analyser bank. The colored area indicate the resolution range for the other analyser banks. The solid lines indicate the results of the analytical calculations. }
    \label{fig:resolution_cuts_comparison}
\end{figure}

The energy-transfer uncertainty is modelled by contributions from the uncertainty in the final neutron energy, $\sigma_{E_{\rm f}}$, arising from the finite angular acceptance of the analyser system in the secondary spectrometer, and from the uncertainty in the time-of-flight, $\sigma_{t_{\rm ToF}}$, which is defined as the opening time of the PSC chopper. 
\begin{equation}
    \delta E = \sigma_{\Delta E} = \sqrt{ \left( \frac{\partial \Delta E}{\partial t_{\rm ToF}} \right)^2 \sigma_{t_{\rm ToF}}^2 + \left( \frac{\partial \Delta E}{\partial E_{\rm f}} \right)^2 \sigma_{\rm E_{\rm f}}^2}
    \label{eq:e_resolution}
\end{equation}
Several potential contributions to the energy resolution are not included in the present treatment. Effects due to finite detector depth and uncertainties in neutron flight-path lengths are neglected, as their associated time-of-flight uncertainties are small compared to the dominant contributions considered here. Contributions arising from analyser crystal lattice-spacing variations, expressed through terms proportional to $(\Delta d)/d$, are likewise not included.
 We present the derivatives separately: 
\begin{eqnarray}
    \left( \frac{\partial \Delta E}{\partial t_{\rm ToF}} \right)^2 &=& \left( \frac{L_{\rm i}^2 m_{\rm n}}{\left(  t_{\rm ToF} -  t_{\rm f}\right)^3}\right)^2 \, ,
    \label{eq:derivative_tof} \\
\left(\frac{\partial \Delta E}{\partial E_{\rm f}}\right)^2 &=& \left(1 + \frac{L_{\rm i} ^2 m_{\rm n} t_{\rm f}}{2E_{\rm f}(t_{\rm ToF}-t_{\rm f})^3} \right)^2 \, .
    \label{eq:derivative_ef}   
\end{eqnarray}

In eq.~(\ref{eq:derivative_ef}), the first term of the right-hand side is the energy resolution contribution from the secondary spectrometer, while the second term describes the correlation between the time-of-flight determination of $E_{\rm i}$ and $E_{\rm f}$, through the secondary spectrometer time-of-flight value $t_{\rm f}$.
Our result is one level of complexity deeper than the simpler approximations we presented in Ref.~\cite{toft-petersen_bifrostindirect_2025}.

The analytical values of $\delta E$ are initially obtained in units of statistical standard deviation (Gaussian $\sigma$). However, in order to comply with experimental practice, we convert $\delta E$ into units of Gaussian full-width-at-half-maximum (FWHM), by multiplying with the factor $2 \sqrt{2 \ln 2} \approx 2.35$. Selected examples are shown in Fig.~\ref{fig:resolution_cuts_comparison}, as solid lines, and the corresponding values are given in table \ref{tab:Analytica_energy}, appendix \ref{ap:analytical_energy}.

To evaluate the energy resolution, with the included terms presented here, a series of McStas simulations was carried out, including a realistic representation of the neutron pulse, beam transport, choppers, and a single of the 9 wedges of the secondary spectrometer with all 5 analyser energies. We do not include the detector depth nor the combined beryllium filter and collimator in the model.
Time focusing was applied from the ESS source to efficiently sample only neutrons reaching the pulse shaping choppers in the (partially) open position. The sample was modelled to mimic incoherent scattering with a tuneable energy transfer. This allowed identification of the broadening of the isolated instrumental energy as a function of energy transfer and PSC opening time. Higher-order Bragg scattering from the analysers was suppressed. 

In Fig.~\ref{fig:resolution_cuts_comparison}, we compare the simulated energy resolutions with the analytical resolution on a log-log scale. The simulation results are plotted as filled symbols, while the lines show the analytical predictions. The different colors represent different instrument settings. 
The simulated values strongly agree with the calculated results. All deviations are small, with a maximum of $\approx 7\%$ deviation as show in the relative residual between the simulations and the theoretical models in appendix \ref{ap:energy_residuals}. The deviations are mostly caused by statistical fluctuations from the simulations. This quantitative agreement gives significant confidence in the validity of the BIFROST McStas Model. However, these results are limited in scope since they only shows the instrument performance in the center of the pulse, while in the experiments, the whole pulse is utilized.

\section{Edge enhancement of energy resolution}\label{sect:edge_enhancement}
\begin{figure}[ht!]
    \centering
    \includegraphics[width=\linewidth]{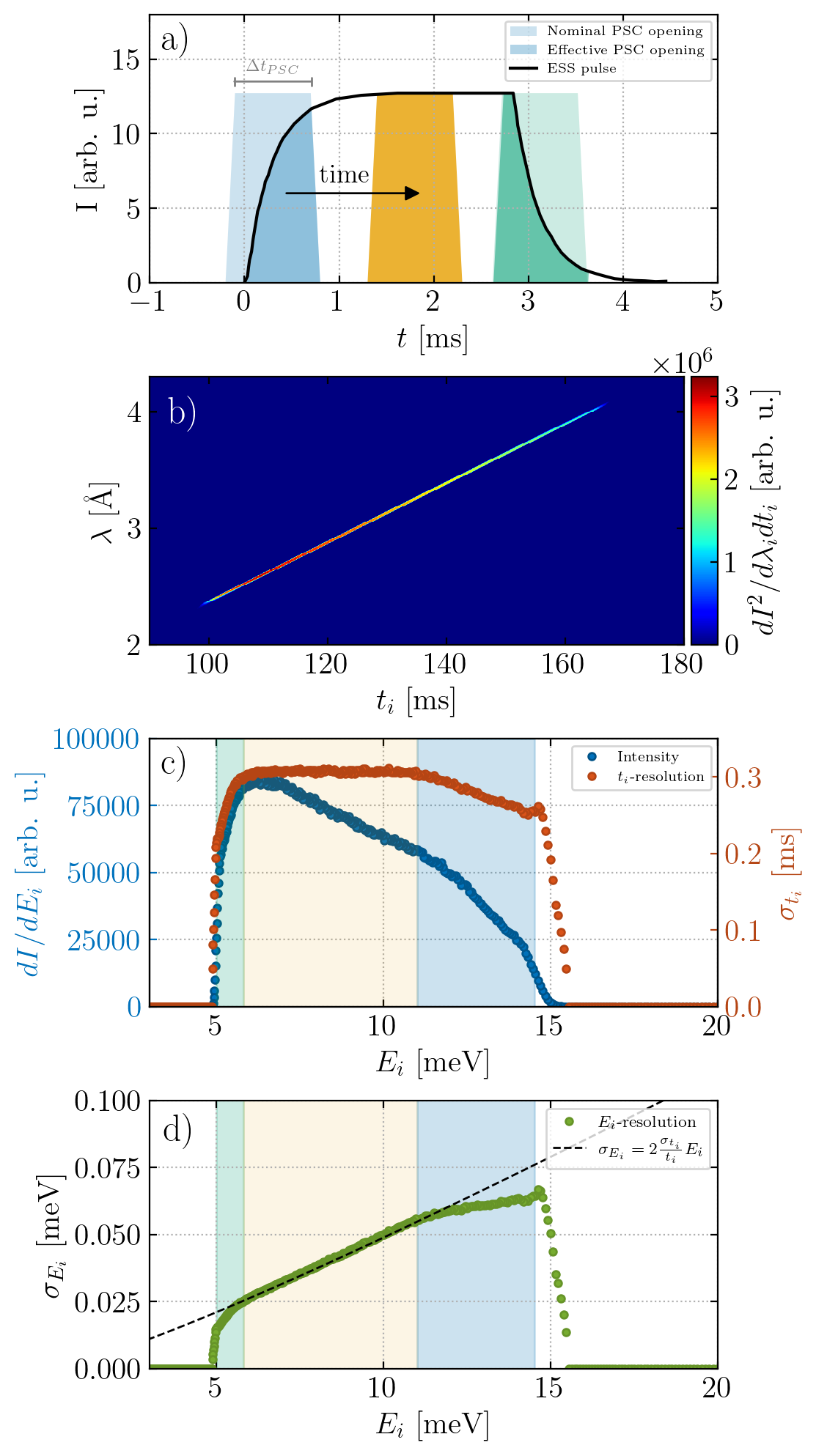}
    \caption{(a) Illustration of the edge enhancement effect, showing how the PSC opening time scans through the moderator pulse as the neutron arrival time changes. The PSC opening will only be partly covered in the beginning (blue) and the end (green) of the pulse. (b) Simulation with $\Delta t_{\rm PSC} = 0.3$~ms, showing the correlation between true neutron wavelength, $\lambda$, and the actual $\sigma_{t_i}$ at the sample position. The wavelength is converted to energy in (c), which shows $\sigma_{t_i}$ vs. initial energy (red points), and total neutron intensity vs. initial energy across the band (blue points). The orange area show the energies where $\sigma_{t_i}=\Delta t_{PSC}$, while the green and blue areas show the energies of the edges effects, which are identified by a decrease in $\sigma_{t_i}$. The same areas are plotted on (d) where $\sigma_{t_i}$ is converted to $\sigma_{E_i}$ vs.\ $E_i$ and the expected behaviour from eq.~(\ref{eq:ei_ti}) is shown by the black dashed line.}
    \label{fig:t_vs_lambda}
\end{figure}

Our calculations in the previous section assume that the PSC pulse width is constant in time.
However, due to the pulse structure of the ESS moderator, this is not always the case.
What is not covered by our method is a particular effect which arises towards the edges of the wavelength band, due to how the PSC opening ($\Delta t_{PSC}$) views the moderator pulse. As shown in Fig.~\ref{fig:t_vs_lambda}(a), the PSC opening is scanning through the moderator pulse as the arrival time changes. In the beginning (blue) and the end (green) of the pulse, only a part of the PSC opening is covered by the long pulse, leading to an artificially smaller $\sigma_{t_i}$, which in turn improves the resolution of the incoming neutron energy. In contrast, in the central part of the pulse (orange), the full PSC opening time is filled and thus $\sigma_{t_i} = \Delta t_{\rm PSC}$. In this section, we investigate this {\em edge enhancement effect} and discuss its significance for real experiments.

To illustrate this effect, we simulate a full wavelength band with the bottom of the band set to $\lambda_0 = 2.3$ Å. Fig.~\ref{fig:t_vs_lambda}(b) shows the relation between the true neutron wavelength, $\lambda_i$, and the measured time-of-flight at the sample position, $t_i$. We see the expected linear correlation between $t$ and $\lambda_i$, while the ''thickness`` in either direction of the line represents the uncertainty of $t$ and $\lambda$. We plot the $t$-uncertainty along with the neutron intensity as a function of $E_i$ as a line plot in  Fig.~\ref{fig:t_vs_lambda}(c). 
Here we see, in the central orange part of the panel, that $\sigma_{t_i}$ is in general constant across the central part of $E_i$, as expected. However, towards the ends of the band we observe that $\sigma_{t_i}$ decreases significantly. The regions of this edge effect are marked with blue and green shading in Fig.~\ref{fig:t_vs_lambda}(c) and hence show the regions of $E_i$, where the resolution deviates from our analytical expectations discussed previously. This resolution improvement comes with a cost of neutron intensity as also seen in Fig.~\ref{fig:t_vs_lambda}(c). 

To facilitate a clear interpretation of our results, Fig.~\ref{fig:t_vs_lambda}(d) shows the resulting conversion of $\sigma_{t_i}$ into $\sigma_{E_i}$ via error propagation;

\begin{equation}
    \sigma_{E_i} = 2 \frac{\sigma_{t_i}}{t_i}E_i .
    \label{eq:ei_ti}
\end{equation}

Here, in the edge-enhanced regions, we observe how the energy resolution is lower than the expected behaviour following eq.~(\ref{eq:ei_ti}), assuming $\sigma_{t_i} = \Delta t_{\rm PSC}$, as shown with a dashed line.

Another consequence of the edge enhancement effect is that the time asymmetry of the pulse will affect the otherwise linear relation between true wavelength and average time-of-flight at the ends of the wavelength band and must be taken into account in the instrument calibration procedures.

This edge enhancement effect can be regarded as an intrinsic enhancement of the energy resolution, as it originates from the instrument itself and will therefore always be present in experimental data. Consequently, when selecting measurement configurations, it is important to identify where these regions occur, since they can influence the interpretations of the instrument resolution during data analysis.  The presence of these regions also offers additional opportunities for experimental design, enabling instrument settings to be chosen such that areas of improved energy resolution coincide with features of interest in the spectrum. As illustrated by the large region at high $E_i$, this may be particularly advantageous for studies at higher energy transfers. A detailed treatment of how to incorporate this behaviour into a complete analysis is left for future work.

\section{Q-resolution}

\begin{figure*}[t]
    \centering
    \includegraphics[width=\linewidth]{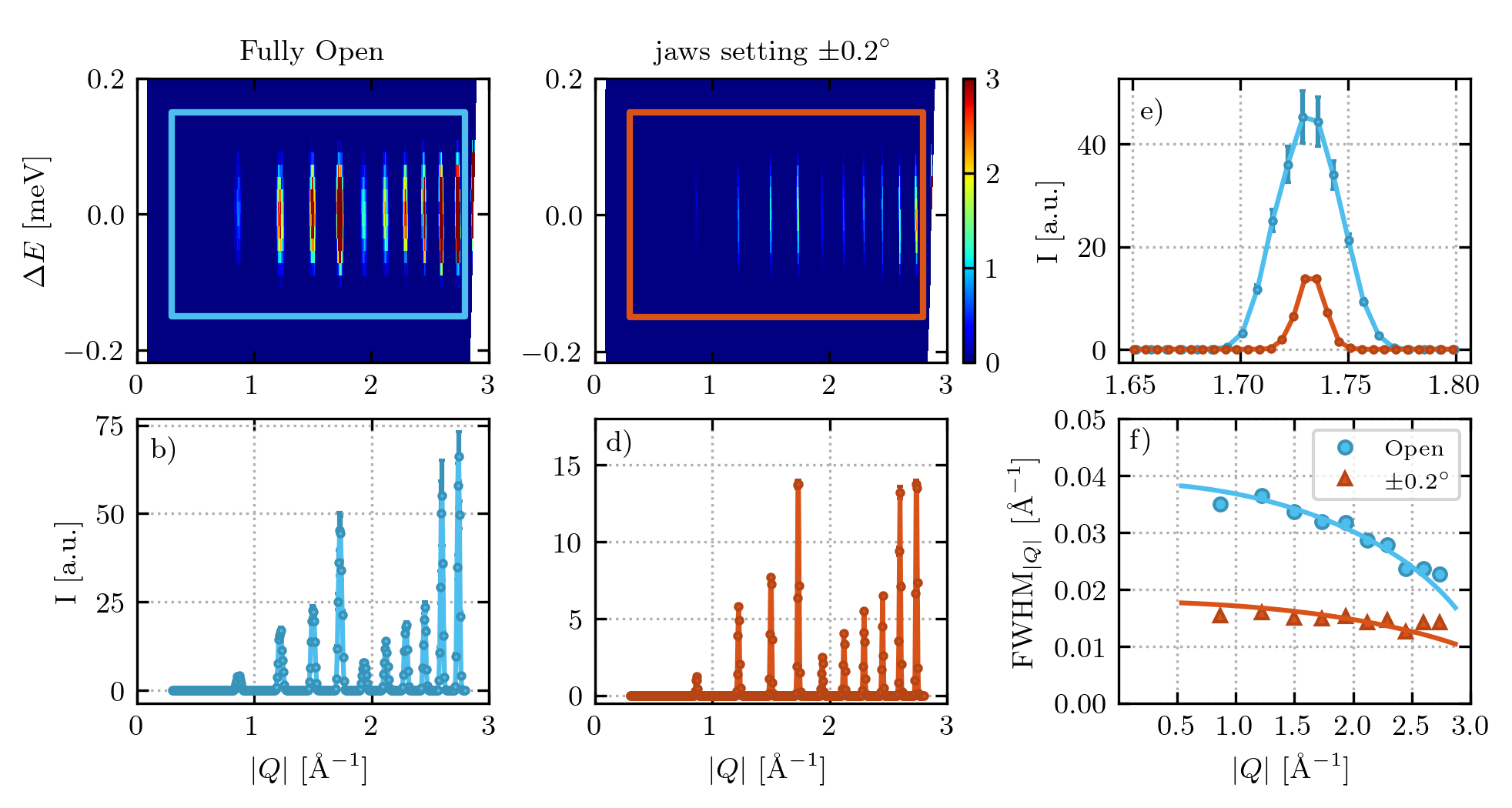}
    \caption{Simulated NaCaAlF powder peaks for fully open (a,b) and $\pm 0.2 ^{\circ}$ (c,d) divergence-jaw settings. Panels (a,c) show the raw $(|Q|,\Delta E)$ distributions, while panels (b,d) display the corresponding energy-integrated profiles $\pm 0.15$~meV. Panel (e) compares representative peaks, and panel (f) summarizes the resulting FWHM values from the simulations (dots) with our theoretical predictions (dashed lines), illustrating the improved $Q$-resolution obtained with tighter jaws from both simulations and analytical calculations.}
    \label{fig:Q_resolution}
\end{figure*}

Having examined the influence of instrument parameters on the energy resolution, we now turn to the corresponding effects on the momentum-transfer resolution. The pure $Q$-resolution is best tested via a powder scan, and therefore we outline the definitions of the in-plane analytical expressions and use them to estimate the resolution of $|Q|$. We find the components of $Q$ via the expression,

\begin{equation}
    {\bf Q} = {\bf k}_i-{\bf k}_f = |k_i| \begin{pmatrix} \cos(\gamma_i) \\ \sin(\gamma_i) \\ 0 \end{pmatrix} - |k_f| \begin{pmatrix} \cos(\phi) \\ \sin(\phi) \\ 0 \end{pmatrix}.
\end{equation}

Here $Q_x$ is the component parallel to the incident beam, $Q_y$ is the $Q$-component perpendicular to the incident beam, and we ignore the small vertical component of momentum transfer, $Q_z$. The different parameters are; $\phi$ is the scattering angle from the sample, $\gamma_i$ is the incident horizontal divergence on the sample, $\theta_A$ is the take-off angle from the analyser and $|k_i|$ ($|k_f|$) are the nominal magnitudes of incident (final) wave vectors, respectively. We include these effects in order to get an accurate description of the effect from the four divergence jaws, which controls the incoming horizontal divergence on the sample.

The variance of the two relevant components of the momentum transfer is found by error propagation; here displaying only the leading terms:
\begin{equation}
    \delta Q_x = \sqrt{\left( \frac{\partial Q_x}{\partial t_{\rm ToF}} \right)^2 \sigma_{t_{\rm ToF}}^2 + \left(\frac{\partial Q_x}{\partial \phi}\right)^2\sigma_{\phi}^2+\left(\frac{\partial Q_x}{\partial \theta_{\rm A}}\right)^2\sigma_{\theta_A}^2}
    \label{eq:qx_resolution}
\end{equation}
and
\begin{equation}
    \delta Q_y = \sqrt{ \left(\frac{\partial Q_y}{\partial \gamma_i}\right)^2 \sigma_{\gamma_{\rm i}}^2 +  \left(\frac{\partial Q_y}{\partial \phi}\right)^2\sigma_{\phi}^2+ \left(\frac{\partial Q_y}{\partial \theta_{\rm A}}\right)^2 \sigma_{\theta_{\rm A}}^2}
    \label{eq:qy_resolution}
\end{equation}

These expressions are used in the error propagated expression of $|Q|= \sqrt{Q_x^2+Q_y^2}$ which are written out in appendix \ref{ap:Q_analytical}, together with the different terms of the derivatives in eqs.~(\ref{eq:qx_resolution}) and (\ref{eq:qy_resolution}).

To assess how these analytically derived contributions manifest in practice, we can again perform a virtual experiment. For this purpose, it is convenient to use a powder sample with many reflections. We have chosen the material Na$_2$Ca$_3$Al$_2$F$_{14}$ (NaCaAlF), which exhibits many closely spaced (but distinct) reflections. NaCaAlF is often used for calibration and is well tested in McStas \cite{Willendrup2006}.

The simulations were performed using the \texttt{PowderN}-component where the sample structural information is input from the McStas system file \texttt{Na2Ca3Al2F14.laz}. Measurements were performed over 4 settings in tank offsets of $6^{\circ}, 11^{\circ}, 47^{\circ}$ and $52^{\circ}$ relative to the direct beam, in order to cover the fully available range of $|Q|$. Furthermore, we test how to improve the Q-resolution by tuning the divergence jaws. We compare a fully open setting to a tight setting of the jaws, giving an incoming divergence of $\pm 0.2^{\circ}$. 

The raw simulation results are shown in Fig.~\ref{fig:Q_resolution}, panels (a) and (c). To analyse the peak widths, we integrate over the energy resolution window $\Delta E = \pm 0.15$~meV, producing the one-dimensional profiles displayed in panels (b) and (d). These profiles were subsequently fitted with Gaussian functions. The full numerically determined peak widths are provided in the appendix.

Fig.~\ref{fig:Q_resolution}(e) presents a direct comparison of two corresponding peaks, clearly illustrating that the tighter divergence-jaw settings yield a substantially improved $Q$-resolution. The full width at half maximum (FWHM) of the powder peaks as a function of $|Q|$ is shown in Fig.~\ref{fig:Q_resolution}(f), and the analytical prediction from the above-mentioned calculations are shown with dashed lines, showing excellent agreement with the behaviour from the simulations. From these results, we can see how the resolution improves from 0.03 Å$^{-1}$ to 0.015 Å$^{-1}$ at $|Q|= 2.0$ Å$^{-1}$. However, this comes at a cost, since the measured peak intensity is reduced by a factor of 3.

Although BIFROST is not designed for powder diffraction, these simple simulations reproduce both the peak positions of the NaCaAlF sample and the expected $Q$-resolution. These results provide strong validation of the modelling setup and further reinforce confidence in the accuracy of our approach.

\section{Resolution of dispersive modes}

\begin{figure*}[t]
    \centering
    \includegraphics[width=\textwidth]{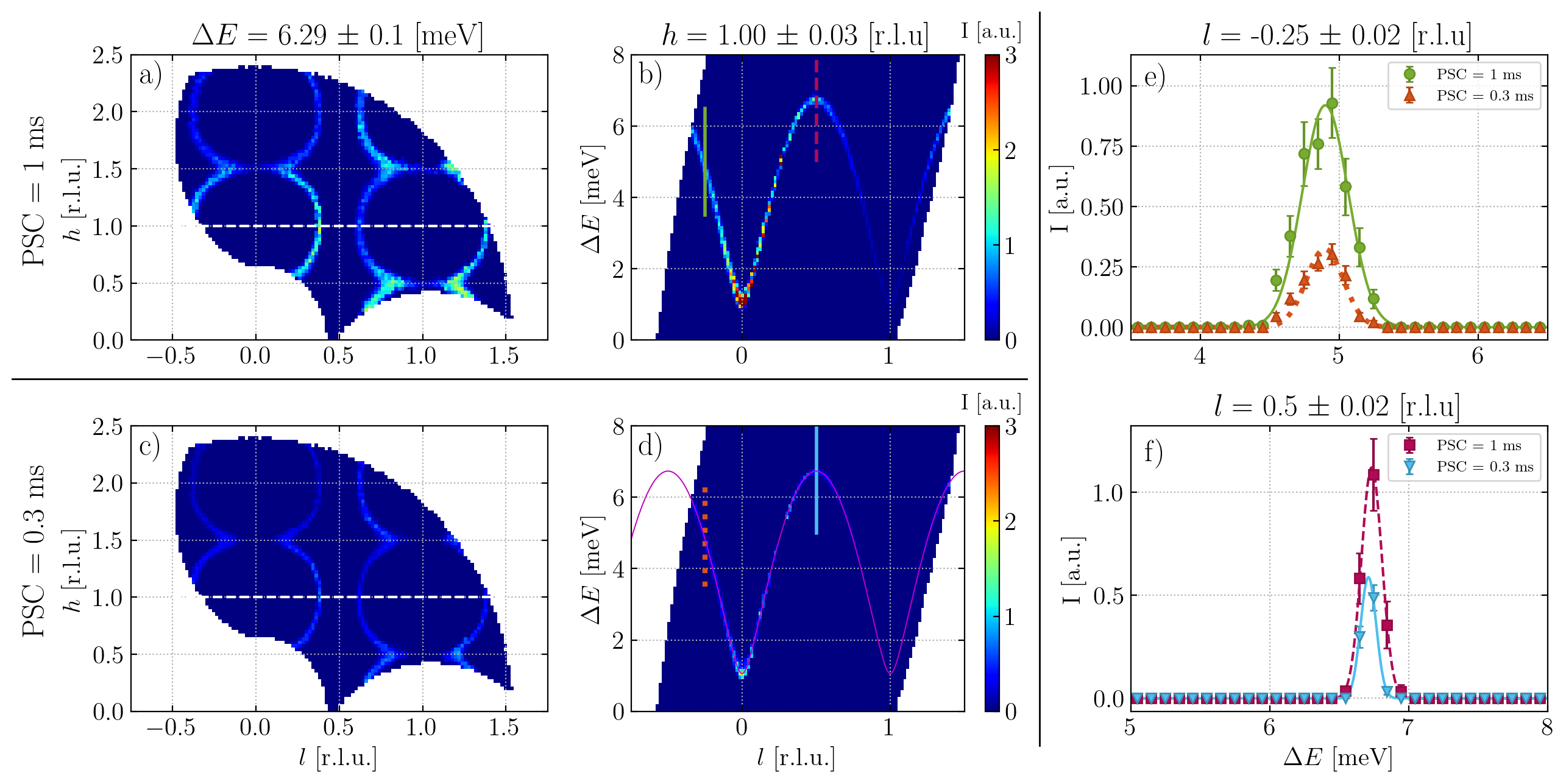}
    \caption{Simulated spin wave dispersion relations for MnF$_2$ for 2 different PSC settings; 1.0~ms (a-b) and 0.3~ms (c-d). The dashed lines in (a) and (c) show the cut direction of constant $h=1$~r.l.u.\ for the q, E-plots (b) and (d). In (d) the theoretical dispersion MnF$_2$ is overlaid with the solid purple line. Constant q-cuts compare the signal intensity and width between PSC chopper settings in (e) for $l=-0.3$ and in (f) for $l=0.5$.}
    \label{fig:MnF2dispersion}
\end{figure*}

Having tested both the energy and the momentum transfer, we now test for a typical use case of the spectrometer: A dispersion relation. Therefore, we will now present simulation results of a dispersive mode in BIFROST at different instrument settings. In this way, we demonstrate the resolving power of the instrument as well as gain insight into the correlation between $Q$- and energy-resolution.

We simulated the spin wave signal of the well-known antiferromagnet (AF) MnF$_2$, thoroughly described {\em e.g.}\ in Ref.~\onlinecite{yamani_neutron_2010}.
We use the new McStas component \verb|SpinWave_BCO|, described in Ref.~\cite{schack_accurate_2025} and apply the same parameters as given in the reference. For the simulation, we use the wavelength band where $\lambda_0 = 2.4 $~\AA\ and here present only the data from the 5.0~meV analysers. This means that we will observe the edge effect in the range $\Delta E =$ 0-1~meV. However, this will not be significant, since the dispersion has a gap of $\approx$ 0.8~meV. Following the observation in section \ref{sect:edge_enhancement}, we expect the edge effect in the top of the energy transfer region to set in at $\Delta E\approx 6$~meV.  We perform the simulations by having the crystal ($a,c$)-plane in the BIFROST scattering plane, and having the c-axis parallel to the incoming beam. We rotate the MnF$_2$ sample $\omega = [-26^\circ ; 76^\circ]$ in $1^\circ$ steps. The sample rotation is performed for two tank settings, $\phi = [15^\circ, 20 ^\circ]$, to cover dark angles.

The simulation results of two PSC settings, but otherwise identical scans, are shown as cuts in the 3D $(h, l, \Delta E)$ space in Fig.~\ref{fig:MnF2dispersion}. Here, (a, b) show the PSC setting of 1.0~ms, where (a) is $(Q, Q)$-cut and (b) is a $(Q, E)$-cut. Likewise, the two panels Fig.~\ref{fig:MnF2dispersion}(c, d) show the similar cuts with the PSC setting of 0.3~ms. The expected dispersion relation of MnF$_2$, described elsewhere\cite{schack_accurate_2025}, is overlaid in (d) where we see an excellent agreement between expectations and the simulated data. 
Some residual slanting of the dispersion is visible in panels (b) and (d), where the positive-slope branch is narrower than the negative-slope branch. Furthermore, the positive-slope branch is less densely sampled, as the dispersion becomes locally parallel to the McStas sampling trajectory, which reduces the sampling probability in these regions.

Fig.~\ref{fig:MnF2dispersion}(e) compares constant-$Q$ cuts of the two PSC settings at $l=-0.3$ while (f) for $l=0.5$. The FWHM from the Gaussian fits from the different cuts and their corresponding configuration are listed in Tab.~\ref{tab:resolution comparison}. For both values of $l$, we are able to see several interesting features. 
First of all, we see how the decrease in PSC opening time improves the energy resolution, at the expense of intensity. 
Furthermore, we see how the energy resolution varies between the different $Q$-cuts. Fig.~\ref{fig:MnF2dispersion}(e) shows a broadening caused by the slope of the dispersion, due to the ($\pm 0.02$ r.l.u.) integration width in $l$, making the apparent energy resolution broader than the raw energy resolutions in Fig.~\ref{fig:resolution_cuts_comparison}. In contrast, Fig.~\ref{fig:MnF2dispersion}(f) shows a cut on the top of the dispersion, resulting in significantly sharper energy resolution, as the slope of the dispersion here is zero. In fact, when comparing the $E_f = 5.0$~meV, $PSC=1$~ms at $\Delta E = 7$~meV in table \ref{tab:resolution comparison} to the results from the idealized energy resolution results from section \ref{sect:energy_resolution}, given in table \ref{tab:Analytica_energy}, the values show excellent agreement.

\begin{table}[h!]
\centering
\begin{tabular}{c | c | c || c}
\hline
PSC-$\Delta t$ & $l$-cut [r.l.u] & $\Delta E$ [meV]  & FWHM$_{\Delta E}$ [meV]\\
\hline\hline
1~ms & -0.25 & 4.8 & 0.39 \\
0.3~ms & -0.25 & 4.8 & 0.31 \\
1~ms & 0.5 & 6.7 & 0.17 \\
0.3~ms & 0.5 & 6.7 & 0.13 \\
\hline
\end{tabular}
\caption{Fitted FWHM$_{\Delta E}$ values (in~meV) for two pulse shaping chopper opening times (PSC–$\Delta t$) at selected $l$-cut positions, given in reciprocal lattice units.}
\label{tab:resolution comparison}
\end{table}

Taken together, these results demonstrate how the simulated commissioning experiment provides a coherent picture of the interplay between the instrument configuration, the intrinsic instrument resolution, and the characteristics of the underlying excitation. The comparison of PSC settings confirms the expected trade-off between resolution and intensity, while the constant-$Q$ cuts illustrate how the local slope of the dispersion governs the apparent broadening relative to the intrinsic energy resolution. The agreement between the simulated peak widths and independently estimated values reinforces the validity of both the analytical description and the virtual-experiment approach. Overall, the MnF$_2$ case demonstrates the level of detail that can be anticipated throughout instrument design and commissioning. This underscores the broader value of such simulations in benchmarking expected performance and in selecting appropriate experimental settings for future measurements.

\begin{figure*}[ht]
    \centering
    \includegraphics[width=\textwidth]{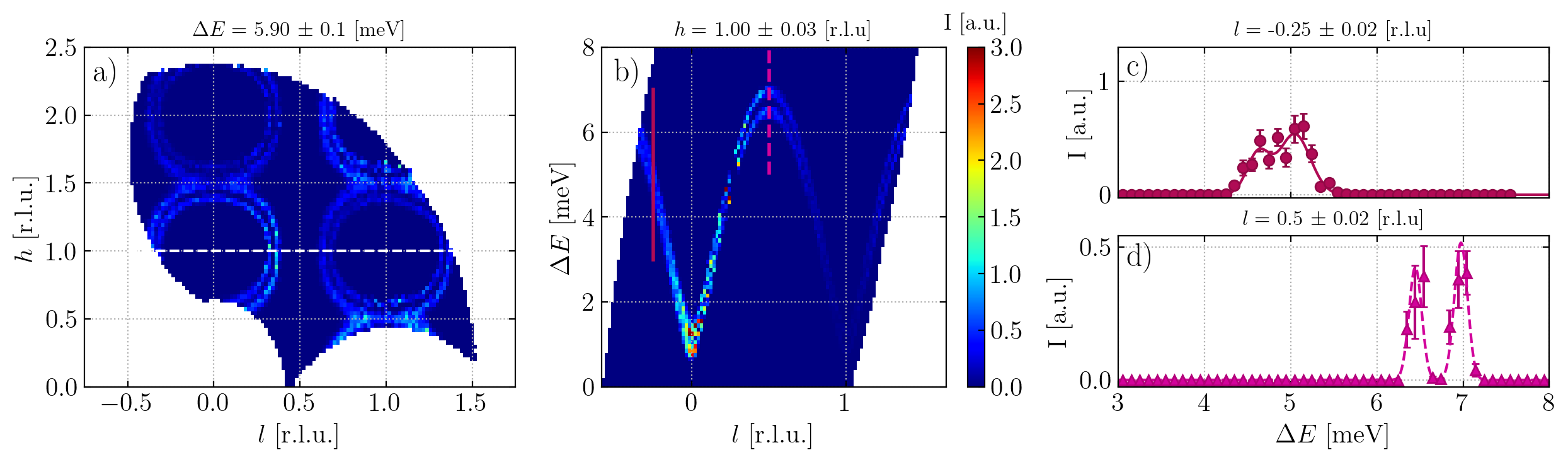}
    \caption{Simulated spin wave dispersion relations for MnF$_2$ with PSC$ = 1$~ms and Zeeman splitting induced by a 2 T field; The dashed line in (a) show the cut direction of constant $h=1$ r.l.u. for the $Q$, E-plot at (b). Constant $Q$-cuts on (c) compare the signal intensity and branch splitting for $l=-0.25$.}
    \label{fig:MnF2Zeeman}
\end{figure*}

\subsection{Resolution test by Zeeman splitting}

So far, we have demonstrated a simple case of an AF dispersion with a single mode. However, in practice BIFROST will be used for mapping out more complex dispersions, which are rich in different features, such as crystal field level excitations, multi-mode magnons, diffusive spin wave continua, anisotropy gaps, avoided crossings between magnons and phonons, {\em etc}. Our ability to distinguish the physical origins of these phenomena strongly depends on the resolving power of the spectrometer.

In order to demonstrate the resolving power of BIFROST, we have applied a magnetic field to the AF MnF$_2$ sample. This will split the two otherwise degenerate AF spin wave bands and allow us to track at which field values the two Zeeman branches can be distinguished. Similar scans as performed in the previous section are performed for $B = 1, 1.5, 2, 3$ and $5$ T. 
An example dataset for $B = 2$ T is seen in Fig.~\ref{fig:MnF2Zeeman}, while the remaining data are presented in Appendix \ref{ap:B-field}.  In Fig.~\ref{fig:MnF2Zeeman} the data is shown in the same cuts as Fig.~\ref{fig:MnF2dispersion}.

\begin{figure}
     \centering
    \includegraphics[width=\linewidth]{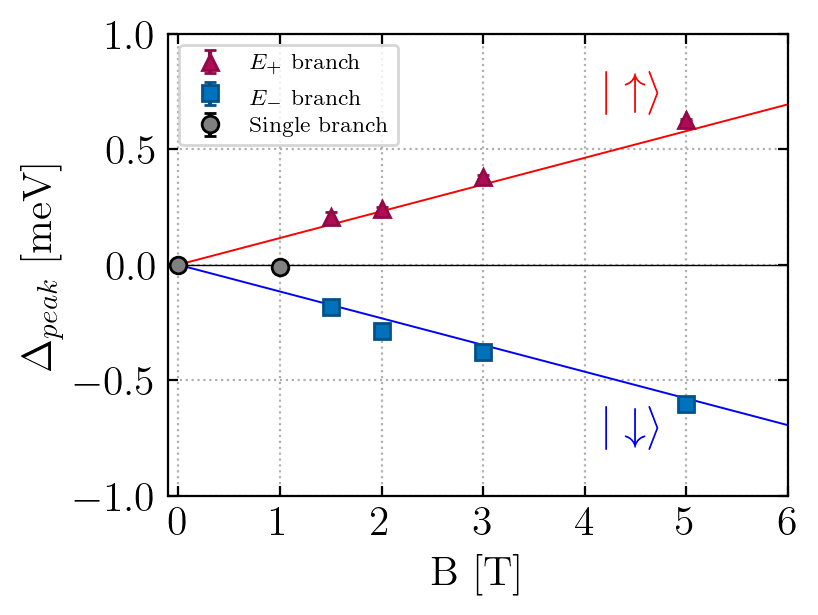} 
    \caption{Zeeman splitting of the two magnon modes at the zone boundary at $l = 0.5$ as a function of the applied B-field. The data show the centers of the peak relative to the zero-field position. The solid red and blue curves show the theoretical expectations for the spin-up and spin-down branches, respectively. Grey points indicate data fitted to a single (unsplit) mode, while the red and blue points correspond to the experimentally resolved split branches. This demonstrates a resolving power of approximately $0.4$~meV for a high flux setting at $\Delta E = 7$~meV.}
    \label{fig:zeeman_dependence}
\end{figure}

We compare the simulated peak splitting with the expected Zeeman energy separation, given by $\Delta_{peak} = 2 g \mu_B B$, as shown in Fig.~\ref{fig:zeeman_dependence}. The analysis focuses on the zone boundary at $l = 0.5$ r.l.u., using the fitted peak position at $B = 0$ T as the reference. For each field value, the fitted peak positions yield the corresponding relative energy shifts. The grey points indicate field values for which two branches could not be separately resolved in the fitting procedure, resulting in a single fitted peak. This illustrates the transition from an unresolved to a resolvable splitting as the field increases.

A comparison of the two cut positions in $l$ shows a clear difference of the peak width pending on the local slope of the dispersion. At the zone boundary at $l=0.5$ r.l.u., where the dispersion is essentially flat, the two Zeeman-split branches are resolved across a wide range of magnetic fields. In contrast, along the sloped region, the peak separation is more obscured by geometric broadening that arises from the finite integration width in $Q$. This difference becomes particularly apparent at intermediate fields, notably at $B= 1.5$ T and $2$ T. This contrast demonstrates how the combination of dispersion slope and instrumental resolution affects the detectability of small energy splittings, and emphasizes the importance of selecting good instrument parameters, that minimize broadening in the regions of interest, when resolving subtle spectral features.

Fig.~\ref{fig:MnF2Zeeman}(a) offers additional insight into the resolving power of the instrument. The Zeeman splitting appears as two clearly separated concentric rings across the entire $(h,l)$-plane. This is noteworthy, because triple-axis spectrometers often show a tilted resolution function that merges the two branches over parts of reciprocal space. In contrast, the nearly symmetrical ring separation in Fig.~\ref{fig:MnF2Zeeman}(a) indicates a more isotropic effective resolution, allowing the band splitting to remain visible over a much larger region. However, some residual slanting remains, as seen in Fig.~\ref{fig:MnF2Zeeman}(b). The positive-slope branch of the dispersion is narrower than the negative-slope branch. 

\section{Discussion and conclusion}

In this work, we have constructed virtual experiments of the BIFROST instrument and validated them against analytical calculations. We find close agreement between simulated and calculated values of both the energy and momentum resolution. The simulations further reproduce the key capabilities of BIFROST across different settings, including a quantitative description of the edge-enhanced energy resolution, which would have been difficult to obtain analytically with comparable precision.

We have assessed the resolving power of the instrument, which is central to the scientific reach of BIFROST. The simulated Zeeman splitting demonstrates, in Fig.~\ref{fig:zeeman_dependence} at $B=2$ T, that BIFROST can clearly resolve an energy separation of approximately 0.4~meV up to transfers of 7~meV in an instrument configurations optimized for flux and not for resolution. Importantly, the achievable resolution shown in these simulations is directly relevant for future scientific cases such as the study of altermagnets\cite{faure_altermagnetism_2025}, where small splittings and subtle symmetry-driven features in the excitation spectrum must be distinguished with high precision far away from the elastic line. The simulations also highlight an important practical aspect of BIFROST. Since the instrument is optimised for small samples, the dispersions observed experimentally will be sharper than those obtained from large co-aligned crystals, where unavoidable mosaicity introduces additional broadening. The virtual experiments presented here therefore provide a realistic indication of the spectral clarity that can be expected in actual measurements, particularly for systems where the intrinsic linewidth is narrow.

The present work also points to areas where further development of simulation tools is needed. While some progress has been seen within simulation of dispersive modes in recent years\cite{winn_flexible_2022}, further realistic modelling of magnetic materials will require additional inelastic components in McStas, including more general dispersion relations, multi-magnon scattering, polarization-dependent cross sections, and magnon-phonon coupling. These ingredients are essential for describing many modern material classes, yet they are not fully implemented within current ray-tracing frameworks. Likewise, more comprehensive analytical treatments of resolution for instruments that combine a time-of-flight source with an indirect-geometry analyser system are required. Substantial progress has been made in analytical resolution theory, for example by Popovici \cite{popovici_resolution_1975} and Violini\cite{violini_method_2014}, but these approaches are typically formulated for monochromator-based or direct geometry instruments and do not capture the full hybrid geometry relevant for BIFROST.

Overall, the results obtained from our virtual experiments provide an initial but substantive assessment of the expected performance of BIFROST. Validation against experimental data will be required once measurements become available. However, this will happen only when ESS delivers a sufficient neutron flux to the instruments. Performing McStas simulations prior to commissioning has already generated insights that will be directly beneficial during commissioning and subsequent operation. More broadly, the present study serves as a proof of concept for virtual experiments on neutron instruments and may help to stimulate the design of experimental strategies at BIFROST, as well as provide a template for similar efforts at other instruments and facilities.

\section*{Acknowledgements}
The project was funded by the Danish National Committee for Research Infrastructure (NUFI) through the ESS-Lighthouse Q-MAT and DanScatt. We thank Peter Willendrup, Mads Bertelsen and Jan-Lukas Wynen from the ESS Data Management \& Scientific Computing Division (DMSC) for McStas and scipp support and Jakob Lass from the Paul Scherrer Institute, for discussions on data reduction.

\bibliography{Main.bib}

\clearpage

\onecolumngrid
\appendix

\section{Energy resolution results}\label{ap:analytical_energy}
Tabulated version of the simulated energy resolutions shown on Fig~\ref{fig:resolution_cuts_comparison} in the main text.

\begin{table*}[ht]
    \small
    \begin{tabular}{c|ccccccccccccccccccc}
        \toprule
        \multicolumn{19}{c}{FWHM$_{\Delta E}$ [$\mu eV$]} \\
        \hline
        Setting vs. $\Delta E$ [meV] & 0 & 1 & 2 & 3 & 4 & 5 & 6 & 7 & 8 & 9 & 10 & 15 & 20 & 25 & 30 & 35 & 40 & 45 & 50 \\
        \hline
         E$_f$ = 2.7~meV,  PSC = 0.1~ms & 15  & 15  & 16  & 17  &  17  & 18  & 20  & 21  & 23  & 24  & 26  & 36  &  49.&  64  & 80  & 100  & 117  & 136  & 153  \\
        E$_f = 5.0$~meV, PSC = 0.1~ms & 39 &  40 & 41 & 41 & 42 & 42 & 44 & 45 & 45 & 47 & 48 & 57 & 70 & 83 & 100 & 119 & 138 & 157 & 179 \\
        $E_f =$2.7~meV, PSC = 1~ms & 22 & 32  & 43  & 56 &  70 &  86 & 103 & 121 & 140 & 158 & 176 & 289 &
  416 & 558 & 723 & 936 & 1025 & 1220 &  1484 \\
        $E_f =$5.0~meV, PSC = 1~ms & 54  & 68 &  84 &  97 & 112 & 130 & 150 & 167 & 189 & 212 & 233 & 352 &
  487 & 650 & 807 & 985 & 1178 & 1418 & 1597 \\
        $E_f =$2.7~meV, PSC = Open & 50 & 84 & 112 & 152 & 196 & 243 & 293 & 343 & 400 & 452 & 522 & 851 & 1232 & 1682 & 2110 & 2645 & 3168 & 3563 & 4372 \\
        $E_f =$5.0~meV, PSC = Open & 131 & 170 & 213 & 262 & 316 & 373 & 429 & 484 & 550 & 611 & 678 & 1056 & 1492 & 1852 & 2389 & 2881 & 3409 & 4198, & 4534 \\  
        \hline
    \end{tabular}
    \caption{Tabulated energy resolutions corresponding to the data shown in Fig.~\ref{fig:resolution_cuts_comparison}. The table reports the measured energy resolution at different energy transfers for the various instrument configurations used, including combinations of analyser final energy ($E_f$) and pulse-shaping chopper (PSC) openings. This tabulation provides a direct numerical reference for the trends illustrated in Fig.~\ref{fig:resolution_cuts_comparison}.}
    \label{tab:Analytica_energy}
\end{table*}

\section{Residuals between theoretical and simulated energy resolution}\label{ap:energy_residuals}

\begin{figure}[h!]
    \centering
    \includegraphics[width=0.7\linewidth]{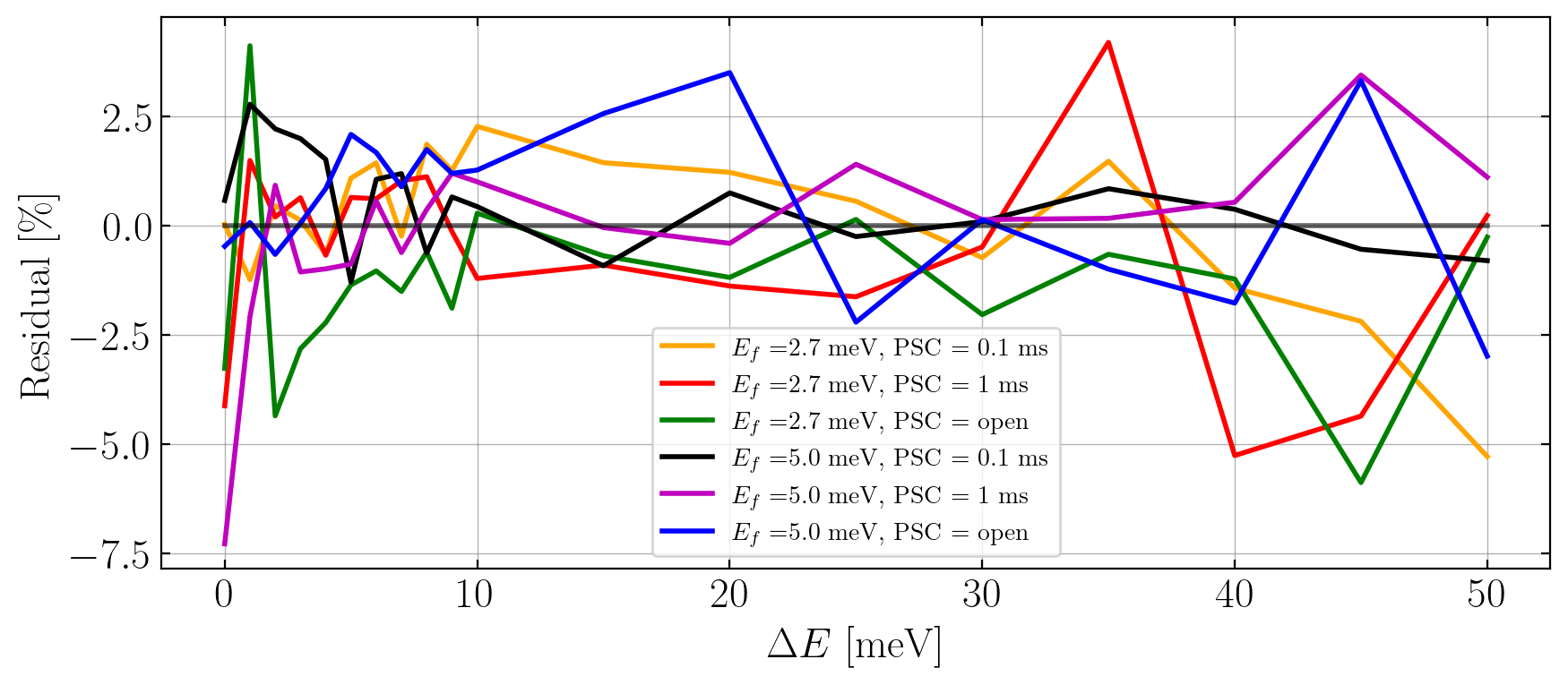}
    \caption{Residual in [\%] between the simulated energy resolution and the analytical prediction given by eq. \ref{eq:e_resolution}.}
    \label{fig:energy_residual}
\end{figure}

\section{Analytical expression of Q-resolution}\label{ap:Q_analytical}
Below, we present a detailed account of the terms described in eqs.~(\ref{eq:qx_resolution}) and (\ref{eq:qy_resolution}).
\begin{equation}
    \left(\frac{\partial Q_x}{\partial \phi}\right)^2 = \left(\frac{|k_{\rm f}|}{t_{\rm ToF} - t_{\rm f}}\right)^2
\end{equation}

\begin{equation}
    \left(\frac{\partial Q_x}{\partial \phi}\right)^2 = |k_{\rm f}|^2\sin^2(\phi)
\end{equation}

\begin{equation}
    \left(\frac{\partial Q_x}{\partial \theta_{\rm A}}\right)^2 = \cot^2(\theta_A)\left( \frac{|k_{\rm i}|t_{\rm f}}{t_{\rm ToF}-t_{\rm f}}+|k_f|\cos(\phi) \right)^2
\end{equation}

\begin{equation}
    \left(\frac{\partial Q_y}{\partial \gamma_i}\right)^2 = |k_{\rm i}|^2
\end{equation}

\begin{equation}
    \left(\frac{\partial Q_y}{\partial \phi}\right)^2 = |k_{\rm f}|^2\cos^2(\phi)
\end{equation}

\begin{equation}
    \left(\frac{\partial Q_y}{\partial \theta_{\rm A}}\right)^2 = |k_{\rm f}|^2\sin^2(\phi)\cot^2(\theta_{\rm A})
\end{equation}

Finally, we present the error propagated expression for the width of $|Q|=\sqrt{Q_x^2+Q_y^2}$.

\begin{equation}
    \delta_{|Q|} = \sqrt{\left(\frac{Q_y}{|Q|}\right)^2\delta Q_y^2+\left(\frac{Q_x}{|Q|}\right)^2\delta Q_x^2+2\frac{Q_xQ_y}{|Q|^2}cov (Q_x,Q_y)^2}
\end{equation}

\section{Q-resolution results}\label{ap:Q_resolution}
In Tab.~\ref{tab:powder_data} below, we present a tabulated version of the Q-resolutions shown in Fig.~\ref{fig:Q_resolution}, of all the individual peaks. 

\begin{table}[h]
\renewcommand{\arraystretch}{1.6}
\centering

\begin{subtable}[t]{0.48\linewidth}
\centering
\begin{tabular}{p{2.5cm} | p{2.5cm} || p{2.5cm}}
\hline
\multicolumn{3}{c}{$Na_2Ca_3Al_2F_{14}$ powder peaks: $\mathbf{ \pm 0.2 ^{\circ}}$ \textbf{Divergence Jaws}} \\
\hline
Peak \# & mean [Å$^{-1}$] & FWHM$_{|Q|}$ [Å$^{-1}$] \\
\hline
1 & 1.2240 & 0.016\\
2 & 1.4997 & 0.014\\
3 & 1.7324 & 0.014\\
4 & 1.9361 & 0.015\\
5 & 2.1208 & 0.014\\
6 & 2.2920 & 0.013\\
7 & 2.4487 & 0.013\\
8 & 2.5974 & 0.014\\
9 & 2.7392 & 0.014\\
\hline
\end{tabular}
\caption{±0.2° divergence jaws}
\end{subtable}
\hfill
\begin{subtable}[t]{0.48\linewidth}
\centering
\begin{tabular}{p{2.5cm} | p{2.5cm} || p{2.5cm}}
\hline
\multicolumn{3}{c}{$Na_2Ca_3Al_2F_{14}$ powder peaks: \textbf{Open Divergence Jaws}} \\
\hline
Peak \# & mean [Å$^{-1}$] & FWHM$_{|Q|}$ [Å$^{-1}$] \\
\hline
1 & 1.2239 & 0.036\\
2 & 1.4999 & 0.035\\
3 & 1.7316 & 0.032\\
4 & 1.9359 & 0.031\\
5 & 2.1211 & 0.029\\
6 & 2.2912 & 0.027\\
7 & 2.4490 & 0.024\\
8 & 2.5972 & 0.024\\
9 & 2.7387 & 0.022\\
\hline
\end{tabular}
\caption{Open divergence jaws}
\end{subtable}

\caption{Tabulated $|Q|$-resolution values corresponding to the data shown in Fig.~\ref{fig:Q_resolution}. 
The tables list the numerically determined powder-peak positions ($\mu$) and full 
widths at half maximum (FWHM$_{|Q|}$) for the nine peaks, read from left to right, 
under two divergence-jaw configurations: $\pm 0.2^\circ$ jaws and fully open jaws.}
\label{tab:powder_data}
\end{table}

\newpage
 
\section{B-field dependence of Zeeman splitting}\label{ap:B-field}
We here show the full data set for the investigation of the field-induced splitting of the AF spin wave modes.\\

\large{\textbf{B = 1 T}}

\begin{figure*}[h!]
    \centering
    \includegraphics[width=0.9\linewidth]{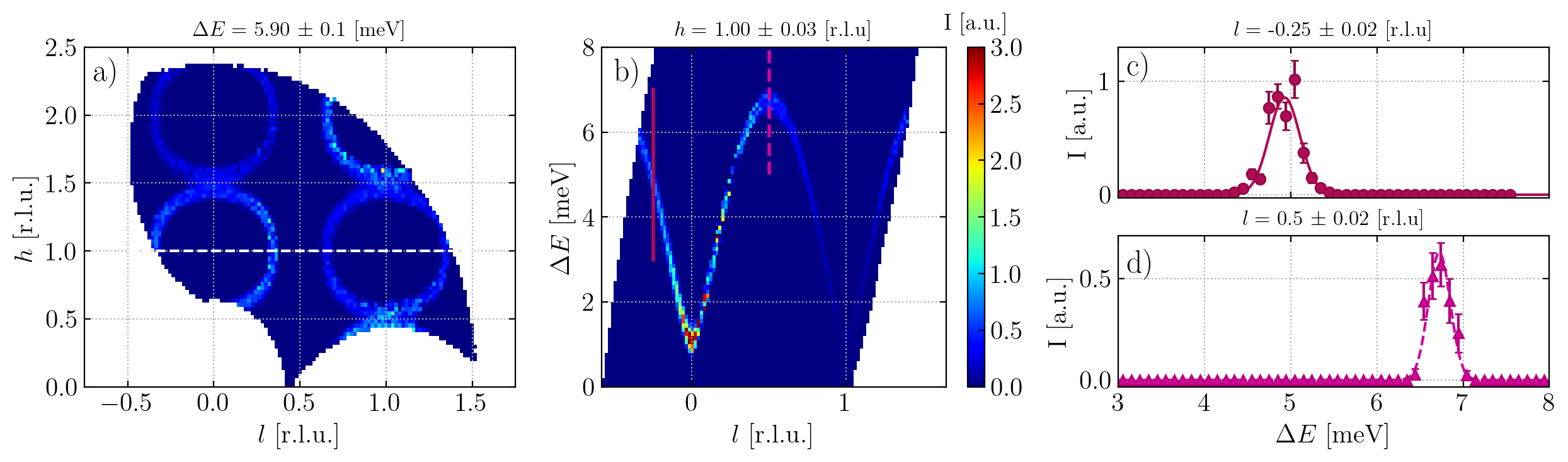}
    \caption{Simulated MnF$_2$ spin-wave dispersion (PSC $= 1$~ms, 1~T). Panel (a) indicates the constant-$h = 1$ cut used in (b); (c) shows constant-$q$ cuts at $l = -0.25$ showing intensity and branch splitting.}
    \label{fig:B-field-1T}
\end{figure*}

\large{\textbf{B = 1.5 T}}
\begin{figure*}[h!]
    \centering
    \includegraphics[width=0.9\linewidth]{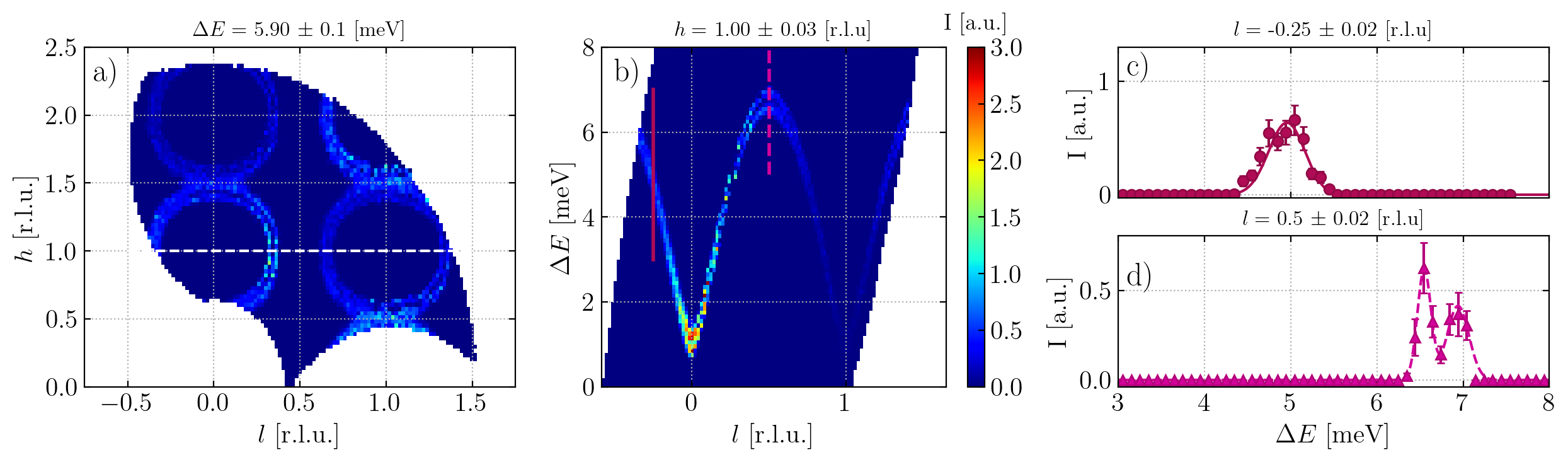}
    \caption{Simulated MnF$_2$ spin-wave dispersion (PSC $= 1$~ms, 1.5~T). Panel (a) indicates the constant-$h = 1$ cut used in (b); (c) shows constant-$q$ cuts at $l = -0.25$ showing intensity and branch splitting.}
    \label{fig:B-field-1p5T}
\end{figure*}

\large{\textbf{B = 3 T}}
\begin{figure*}[h!]
    \centering
    \includegraphics[width=0.9\linewidth]{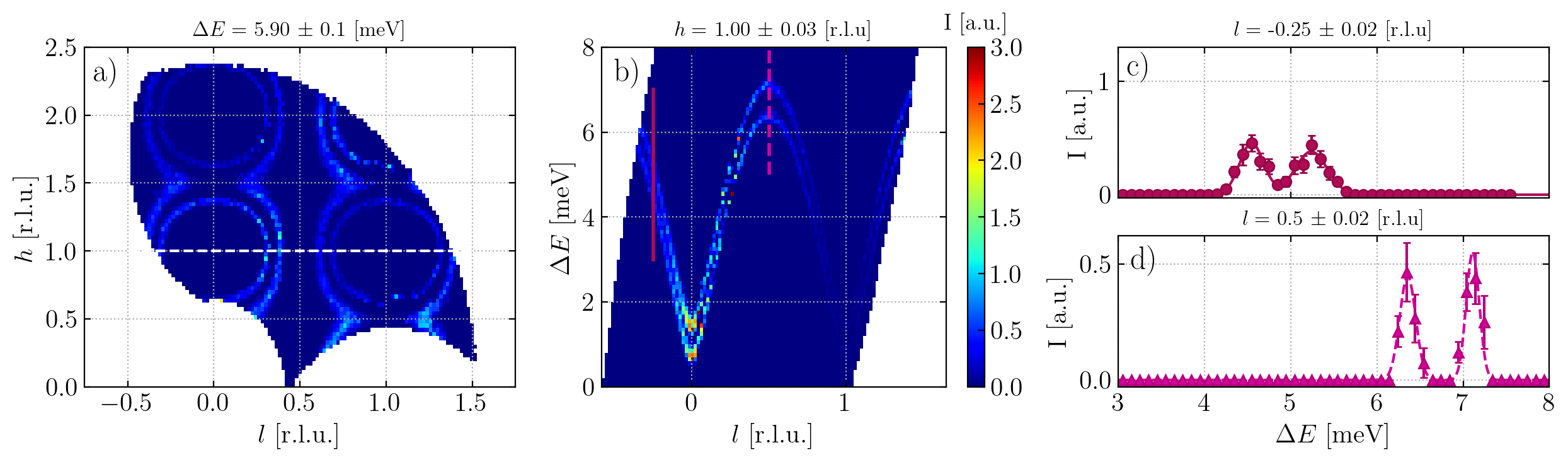}
    \caption{Simulated MnF$_2$ spin-wave dispersion (PSC $= 1$~ms, 3 T). Panel (a) indicates the constant-$h = 1$ cut used in (b); (c) shows constant-$q$ cuts at $l = -0.25$ showing intensity and branch splitting.}
    \label{fig:B-field-3T}
\end{figure*}

\newpage

\large{\textbf{B = 5 T}}
\begin{figure*}[h!]
    \centering
    \includegraphics[width=0.9\linewidth]{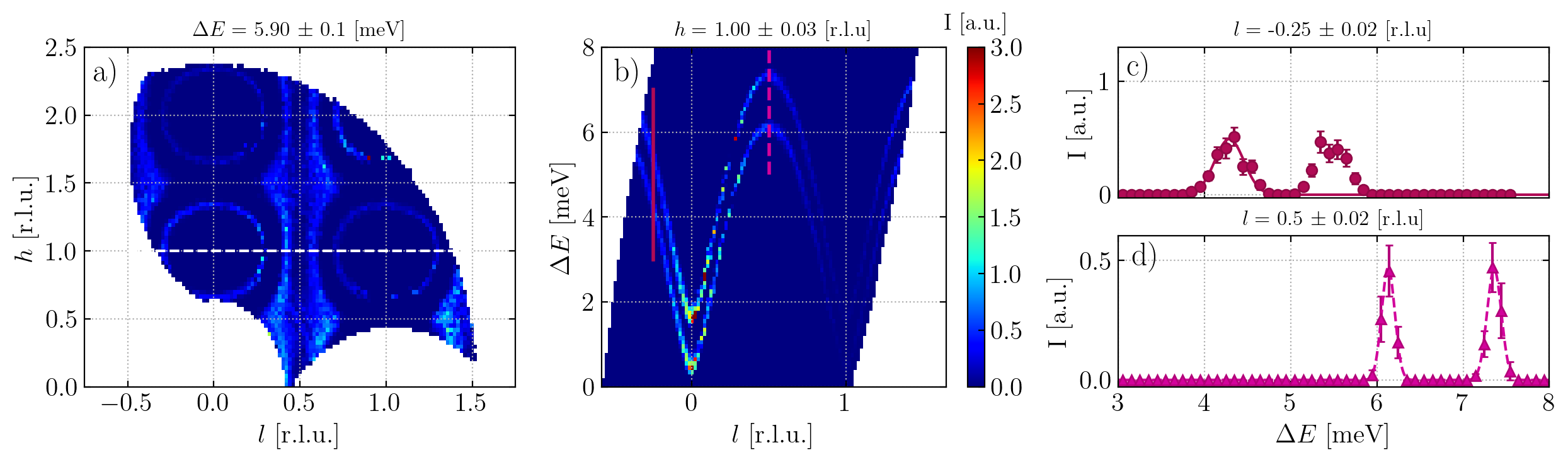}
    \caption{Simulated MnF$_2$ spin-wave dispersion (PSC $= 1$~ms, 5~T). Panel (a) indicates the constant-$h = 1$ cut used in (b); (c) shows constant-$q$ cuts at $l = -0.25$ showing intensity and branch splitting.}
    \label{fig:B-field-5T}
\end{figure*}

\end{document}